\documentclass[prl,twocolumn,amsmath,groupedaddress,amssymb,floatfix,footinbib,bibnotes,longbibliography,superscriptaddress]{revtex4-2}

\usepackage{amsmath,amsfonts,amssymb}
\usepackage{mathrsfs}
\usepackage{graphicx}
\usepackage[english]{babel} 
\usepackage{verbatim}
\usepackage[colorlinks=true,citecolor=blue,linkcolor=magenta]{hyperref}
\usepackage[usenames, dvipsnames]{color}
\usepackage{bbold}
\usepackage{chemformula}
\usepackage{pdfpages}

\makeatletter
\AtBeginDocument{\let\LS@rot\@undefined}
\makeatother

\newcommand{\KITP}{Kavli Institute for Theoretical Physics, University of California, Santa Barbara, CA, 93106-4030}
\newcommand{\UMD}{Condensed Matter Theory Center and Joint Quantum Institute, Department of Physics, University of Maryland, College Park, Maryland 20742-4111}

\begin{document}

\title{Engineering the Kitaev spin liquid in a quantum dot system}

\author{Tessa Cookmeyer}
\email[]{tcookmeyer@kitp.ucsb.edu}
\affiliation{\KITP}

\author{Sankar Das Sarma}
\affiliation{\UMD}
\affiliation{\KITP}

\begin{abstract}
The Kitaev model on a honeycomb lattice may provide a robust topological quantum memory platform, but finding a material that realizes the unique spin liquid phase remains a considerable challenge. We demonstrate that an effective Kitaev Hamiltonian can arise from a half-filled Fermi-Hubbard Hamiltonian where each site can experience a magnetic field in a different direction. As such, we provide a method for realizing the Kitaev spin liquid on a single hexagonal plaquette made up of  twelve quantum dots. 
Despite the small system size, there are clear signatures of the Kitaev spin-liquid ground state, and there is a range of parameters where these signatures are predicted, allowing a potential platform where Kitaev spin-liquid physics can be explored experimentally in quantum dot plaquettes.
\end{abstract}

\maketitle
\section{Introduction}

Quantum spin liquids are a new phase of matter that exhibit the lack of long-ranged order, an emergent gauge field, long-ranged entanglement, topological order, and fractionalization of spins \cite{savary2016quantum, wen2019experimental,broholm2020quantum}. Despite several promising candidate materials coming from frustrated Kagome \cite{feng2017gapped,shores2005structurally} and triangular \cite{Shimizu2003,Itou2008,Law2017,He2018,xu2023realization} lattices, there remains a lack of consensus about the nature of their ground-state phase.

Another route to spin liquid materials comes out of the Kitaev model on the honeycomb lattice \cite{kitaev2006}, an exactly solvable playground for exploring the physics of spin liquids and non-Abelian anyons \cite{Nayak2008}. In the gapless isotropic phase, the low-energy excitations behave as non-Abelian anyons, once a magnetic field introduces a small gap, and these anyons could form the basis for perfect topological memory \cite{Nayak2008}. The model became physically relevant after Jackeli and Khaliullin found that certain materials, arising from 4$d$ rare earth atoms with the correct geometry, may have a significant Kitaev term \cite{Jackeli2009}. 

The search for a material realization of the Kitaev spin liquid, a ``Kitaev material,'' has now generated an enormous amount of research on a host of compounds such as \ch{Na2IrO3} \cite{Ye2012, Comin2012, hwan2015direct,Singh2010,Singh2012,Choi2012,Liu2011}, \ch{Li2IrO3} \cite{Singh2012, Williams2016,Biffin2014, Winter2016}, \ch{H3LiIr2O6} \cite{kitagawa2018,takagi2019}, \ch{Na2Co2TeO6} \cite{lin2021field}, and $\alpha$-\ch{RuCl3} \cite{banerjee2018,Banerjee2017,banerjee2016,Ran2017,nasu2016}. For \ch{RuCl3} in particular, the smoking-gun signature of a Kitaev spin liquid, a quantized thermal Hall effect, has been claimed to have been measured \cite{kasahara2018, yokoi2021, bruin2022}, but convincingly reproducing the results has been difficult and questions remain \cite{yamashita2020,czajka2022,lefranccois2021}. 

Within the Kitaev materials, there remain formidable challenges: most materials enter a long-range ordered phase at low temperature, implying considerable non-Kitaev interactions, and the underlying effective spin Hamiltonian is never known exactly\cite{samarakoon2022,Maksimov2020,Winter2016,Slagle2018,takagi2019}, particularly since the fundamental Hamiltonian is an electronic and not a spin Hamiltonian. In fact, it is unclear that naturally occurring solid state materials can manifest the precise Hamiltonian necessary for producing quantum spin liquids described by theoretical models, including the Kitaev model. 

There is, however, an alternate way of realizing spin liquids by using engineered structures containing the requisite spin Hamiltonian by design, i.e., quantum simulators. Advances in these systems, allow for much more detailed probing of the proposed spin-liquid state. In two-dimensional Rydberg arrays it was theoretically proposed and then  experimentally demonstrated that an arrangement of atoms on the bonds of a Kagome lattice can lead to some topological ordering \cite{Verresen2021,semeghini2021probing}. Although the long-ranged nature of the interaction, non-exactness of the Rydberg blockade, and non-equilibrium nature of the state complicate the interpretation of the experiment \cite{Verresen2021,giudici2022dynamical,sahay2022quantum}, this result represents a definitive advance in spin-liquid experiments. 

Multiple
 proposals for  realizing Kitaev physics in quantum simulators already exist: the first, using ultra-cold atoms \cite{Duan2003}, requires a significant number of independent lasers and two-photon processes and faces serious challenges in cooling the system to low enough temperatures to observe long-ranged topological order \cite{Sun2023Engineering,tarruell2018quantum}. More recent work uses an approach based on a Floquet drive \cite{Sun2023Engineering} and trapped ions \cite{schmied2011quantum}, but the former requires significant temporal coherence and the latter has stringent constraints on the relevant time-scales. There is thus considerable interest in the engineered realization of the Kitaev spin liquid; beyond allowing for the direct access to the physics of spin liquids, topological order and non-Abelian anyons, the braiding of these anyons would allow for quantum computation with ``passive'' quantum error correction \cite{Nayak2008, Terhal2015}. 
 Of course, within fully programmable quantum computers it is possible to directly simulate the Kitaev model \cite{xiao2021determining,bespalova2021quantum} and a similar model with many of the same properties, the toric code \cite{Lu2009,Han2007,xu2023digital}, but these constructions likely benefit less from the topological protection of quantum information. 

In this Letter, we discuss how spin-liquid physics can be explored in small quantum-dot systems by precisely creating the Kitaev model on a single hexagonal plaquette (Fig.~\ref{fig:plaquette}). 
Quantum dot systems are a potential spin qubit quantum computing platform where full control has been demonstrated for six sites \cite{philips2022universal} but where systems with more dots (with as many as sixteen having been fabricated so far \cite{borsoi2023shared}) can be considered quantum simulators of Hubbard-model physics \cite{buterakos2023magnetic}. In fact, semiconductor quantum dot based spin qubits are considered to be a leading quantum computing platform because of their scalability, fast all-electrical operations, and long coherence. Even though the systems are small, they have already provided experimental evidence for Nagaoka ferromagnetism \cite{Nagaoka1966,dehollain2020nagaoka,Buterakos} and the small-system analog of the Mott transition \cite{hensgens2017quantum,Stafford1994}, and they could provide evidence for flat-band ferromagnetism in the near future \cite{Tasaki1992,buterakos2023certain}.  It is already possible to apply a magnetic field gradient using micromagnets \cite{philips2022universal,pioro2007micromagnets}, and our main assumption is that, as the technology improves, it will be possible to place each site in its own effective magnetic field, which is also necessary for quantum computing single qubit operations. Under this assumption, we will derive an effective Hamiltonian that can be tuned to be exactly the Kitaev model on a single hexagon. The physics we propose is no more challenging than fabricating the spin qubit based quantum computing platform, which is a huge activity in more than  a dozen research centers and industrial labs including Intel Corporation \cite{neyens2023probing}.

Since a ``phase'' is only defined in the thermodynamic limit, we cannot claim to ever create a Kitaev spin-liquid ``phase'' in such a small system (which is a problem intrinsic to all quantum simulator platforms). However, we find that the unique properties of the Kitaev model allow for spin-liquid signatures to be manifest even in this small system for a range of parameters around the exact Kitaev point--that is, the system does not have to be perfectly fine-tuned. Our construction is not limited to a single hexagonal plaquette, and can be extended straight-forwardly to a many-unit cell system. 

\section{Theory}

\begin{figure}[tb]
\includegraphics[width=0.45\textwidth]{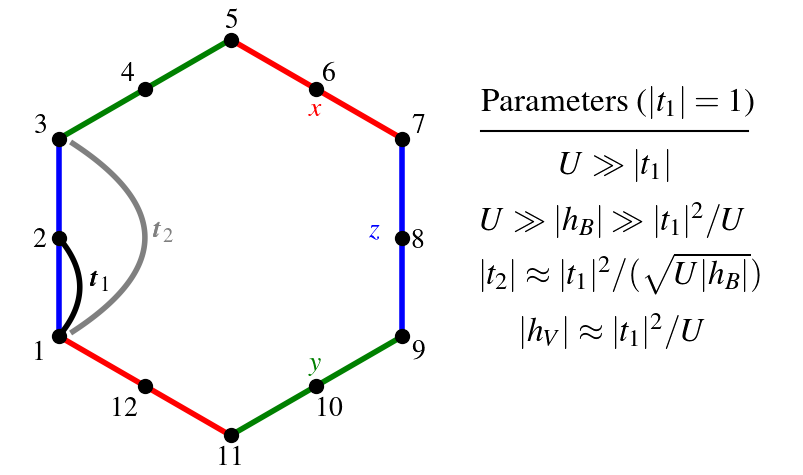}
    \caption{Our system consists of twelve Fermi-Hubbard sites with interaction strength $U$ arranged on a hexagon as shown. We separate the sites into the odd sites which live on the vertices, $V$, and the even sites which live on the bonds, $B$. There is hopping $t_1$ between all adjacent sites and hopping $t_2$ between the vertices.  Even though a next-nearest-neighbor hopping $t_3$ between two bond sites should have roughly the same magnitude as $t_2$, we show in the SM \cite{SM} that its effect is $\mathcal O(U^{-3})$ and does not alter our construction. The six edges of the hexagon are each given a label, $x$, $y$, or $z$ in the pattern indicated by color. The direction of the magnetic field for the bond (vertex) sites points in the direction of the bond label (sum of the two adjacent bond's labels), respectively. For example, $\pmb h_4 = -h_B \hat y$ and $\pmb h_3=h_V(\hat z + \hat y)$. If $|t_1|,|h_B|\ll U$, $|h_B|\gg |t_1|^2/U$, and $|t_2|$ and $|h_V|$ to be related to $|t_1|$, $|h_B|$, and $U$ as shown, then the six vertex spins will interact with an effective Kitaev interaction of strength $K=2|t_1|^4/(U^2|h_B|)$.}
\label{fig:plaquette}
\end{figure}

We start by explaining the construction on a single hexagonal plaquette, see Fig.~\ref{fig:plaquette}. In addition to six sites on the vertices of the hexagon, which will interact via an effective Kitaev Hamiltonian, we have an additional six sites that live on the bonds or edges of the hexagon that will be frozen/integrated out. We will demonstrate that this system, which can be fabricated using the existing spin qubit technology, is sufficient to see Kitaev-spin-liquid-like physics.

Because we are considering an application to experimental quantum dot systems, the Hamiltonian for our twelve-site system is given, by construction, by the Fermi-Hubbard model in a magnetic field 
\begin{equation}
\begin{aligned}\label{eq:12site_Hubb}
    H &=  U\sum_i n_{i\uparrow} n_{i\downarrow} + \sum_{ij,\sigma} t_{ij} c_{i\sigma}^\dagger c_{j\sigma} + \frac12\sum_{i,\sigma,\sigma'} \pmb h_i \cdot c_{i\sigma}^\dagger \pmb \sigma_{\sigma,\sigma'}c_{i\sigma'} 
\end{aligned}
\end{equation}
where $t_{ij}=t_{ji}^*$ are not necessarily real. We assume that the system is half-filled, which is easy to control in spin qubit quantum dot structures.  We only allow for nearest-neighbor hopping $t_1$ and hopping between nearest-neighbor vertices $t_2$, since longer distance hopping falls off exponentially \footnote{Although the hopping $t_3$ between nearest-neighbor bond sites is of the same order as $t_2$, it produces only a small change in the effective magnetic field of the vertex sites at $\mathcal O(U^{-3})$. See the SM \cite{SM}}. We are envisioning two different magnetic field strengths: $|h_B|$ for the sites positioned on the bonds/edges of the hexagon and $|h_V|$ for sites positioned on the vertices. The direction of the magnetic field follows the pattern described in Fig.~\ref{fig:plaquette}: each edge is labeled by one of three orthogonal directions, $x$, $y$, or $z$, and the field on a bond site points in that direction (i.e. $\pmb h_2 = h_B \hat z$), and the field on a vertex points in the sum of the directions of adjacent edges [i.e. $\pmb h_1 = h_V (\hat z + \hat x)$]. We use the hopping strength $|t_1|$ as the energy unit. In order to create a single Kitaev plaquette, we will show self-consistently that we need the scalings $U \gg |h_B| \gg |t_1|^2/U$, $|h_V|\sim |t_1|^2/U$, and $|t_2|\sim |t_1|^2/\sqrt{U|h_B|}$. Again, this is, in principle, achievable in semiconductor  spin qubit platforms, where $U$ and the hoppings are the largest and the samllest energy scales, respectively, and the magnetic field is experimentally tunable.

We first perform perturbation theory in $|h_B|/U, |t_{ij}|/U$ following \cite{takahashi1977half,macdonald1988t}. The full details of the calculation are given in the Supplement Material (SM) \cite{SM},  and we end up with the effective Hamiltonian of localized spins, $\pmb S_i = \pmb \sigma_i/2$, at $\mathcal O(U^{-3})$:
\begin{widetext}
\begin{equation}
\begin{aligned}\label{eq:Hubbpert}
  H_\text{eff,spin} &= \frac{1}{2} \sum_{i} \left(1-\frac{2|t_1|^2}{U^2}\right)\pmb h_i \cdot \pmb \sigma_i+\sum_{\langle ij\rangle} \frac{|t_1|^2}{U} \left(1-4\frac{|t_1|^2}{U^2}+\frac{1}{4} \frac{|h|^2}{U^2}\right)(\pmb \sigma_i \cdot \pmb \sigma_j -1) + \sum_{\langle \langle i i'\rangle \rangle_B} \frac{|t_1|^4}{U^3} (\pmb \sigma_i \cdot \pmb \sigma_{i'} -1) \\
    &+ \sum_{\langle\langle j k\rangle\rangle_V} \left( \frac{|t_2|^2}{U} +\frac{|t_1|^4}{U^3}\right) (\pmb \sigma_j \cdot \pmb \sigma_k -1) + \sum_{\langle ij\rangle} \frac{|t_1|^2}{2U^2} \left( \pmb h_j \cdot \pmb \sigma_i +\pmb h_i \cdot \pmb \sigma_j\right)  +3 \sum_{(j,i,k)_B} \sin(\phi_B) \frac{|t_1^2 t_2|}{U^2} \pmb \sigma_i \cdot (\pmb \sigma_j \times \pmb \sigma_k) \\
\end{aligned}
\end{equation}
where $\langle \langle jk\rangle\rangle_V$ ($\langle \langle i i'\rangle\rangle_B$) indicates next-nearest-neighbor pairs between vertex (bond) sites, and $(j,i,k)_B$ indicates a sum over bonds where $j,k$ are the vertex sites and $i$ is the bond site. The value of $\phi_{B}$ is how much flux, in units of the flux quantum, pierces the triangle made up of those three sites; in our geometry, $\phi_B=0$, but in other geometries this term may exist. However, if $\phi_B \ll 1$, this term is likely negligible. Note that we do not have the ring-exchange term because it requires a 4-cycle to exist in hopping. We also have made use of $|h_V| \sim |t_1|^2/U$ and $|t_2| \sim |t_1|^2/\sqrt{U|h_B|}$ to ignore terms that are already higher-order in $1/U$.

We now integrate out the bond sites to be left with an effective Hamiltonian for just the vertex sites. We fix the magnetic field on each bond to be $\pmb h_i = -|h_B|\hat \alpha$ where site $i$ is on an $\alpha$ bond. We perform perturbation theory in $|t_1|^2/(|h_B|U)$ again to $\mathcal O(U^{-3})$ (with $\phi_B=0$).
\begin{equation}
\begin{aligned}
    &H_\text{V, eff} = \frac12 \sum_{j\in V} \pmb h_\text{eff,$j$} \cdot \pmb \sigma_j + \sum_{\langle jk\rangle_{V,\alpha}} J \pmb \sigma_j \cdot \pmb \sigma_k + K\sigma_j^\alpha \sigma_k^\alpha + C \\
    \pmb h_{\text{eff,$j$}} =  \sum_{i_\alpha \in n.n.(j)}&\hat \alpha \left[ h_V\left(1-\frac{2|t_1|^2}{U^2} \right) + \frac{2|t_1|^2}{U}\left(1-4 \frac{|t_1|^2}{U^2} + \frac{|h_B|^2}{4U^2}-\frac{|h_B|}{2U} + 2\frac{|t_1|^2}{U|h_B|}\right)  \right]\\
    J =\frac{|t_2|^2}{U} + \frac{|t_1|^4}{U^3}-&2 \frac{|t_1|^4}{U^2 |h_B|}+2\frac{|t_1|^4 h_V}{U^2|h_B|^2}-4\frac{|t_1|^6}{U^3|h_B|^2} \qquad K= 2\frac{|t_1|^4}{U^2 |h_B|}-4\frac{|t_1|^4 h_V}{U^2|h_B|^2}+8\frac{|t_1|^6}{U^3|h_B|^2} \\
    \label{eq:vertex_Ham}
\end{aligned}
\end{equation}
\end{widetext}
where $\langle jk\rangle_{V,\alpha}$ indicates nearest-neighbor pairs of vertices that are connected via an $\alpha$ bond, and $i_\alpha \in n.n.(j)$ indicates the nearest-neighbors of site $j$ that are on an $\alpha$ bond. The constant $C$ is provided in the SM \cite{SM}. 

Since we want the field strength, $|\pmb h_\text{eff,$j$}/K \lesssim 1$ and Heisenberg coupling $J/K \lesssim 1$, we need $|t_2| \lesssim \sqrt{2}|t_1|^2/\sqrt{U|h_B|}$ and $h_V \approx -2|t_1|^2/U$, which justifies the scaling we used to derive Eqs.~\eqref{eq:Hubbpert} and \eqref{eq:vertex_Ham}. Although we have computed expressions for these quantities to $\mathcal O(U^{-3})$, we see that the Kitaev coupling is $\mathcal O(U^{-2})$ implying that, even if $\phi_B=0$ turns out to be a poor assumption, our construction still works for large enough $U$. 

Despite the notion of a phase being properly defined only in the thermodynamic limit, there is a clear-cut signature of a Kitaev spin-liquid like ``phase'' even in this small plaquette. First, there is an operator defined on each plaquette that commutes with the Kitaev Hamiltonian. For our single plaquette, it is given by
\begin{equation}\label{eq:plaquette_op}
    W_P =\sigma_1^y \sigma_3^x \sigma_5^z \sigma_7^y \sigma_9^x\sigma_{11}^z =\pm 1
\end{equation}
where the site indices are from Fig.~\ref{fig:plaquette}. The value of $W_P=\pm 1$ is a signature of the emergent $\mathbb Z_2$ gauge field of the Kitaev model \cite{kitaev2006}. Second, the spin-spin correlators are short-ranged \cite{Baskaran2007}: the only non-zero static $S^z$-$S^z$ correlators for our system at the Kitaev point are
    $\langle S_1^z S_3^z\rangle=\langle S_7^z S_9^z\rangle =-1/6$ and $\langle S_j^zS_j^z\rangle = 1/4$ with $\pmb S = \pmb \sigma/2$.

\section{Results}

\begin{figure*}
\includegraphics[width=0.9\textwidth]{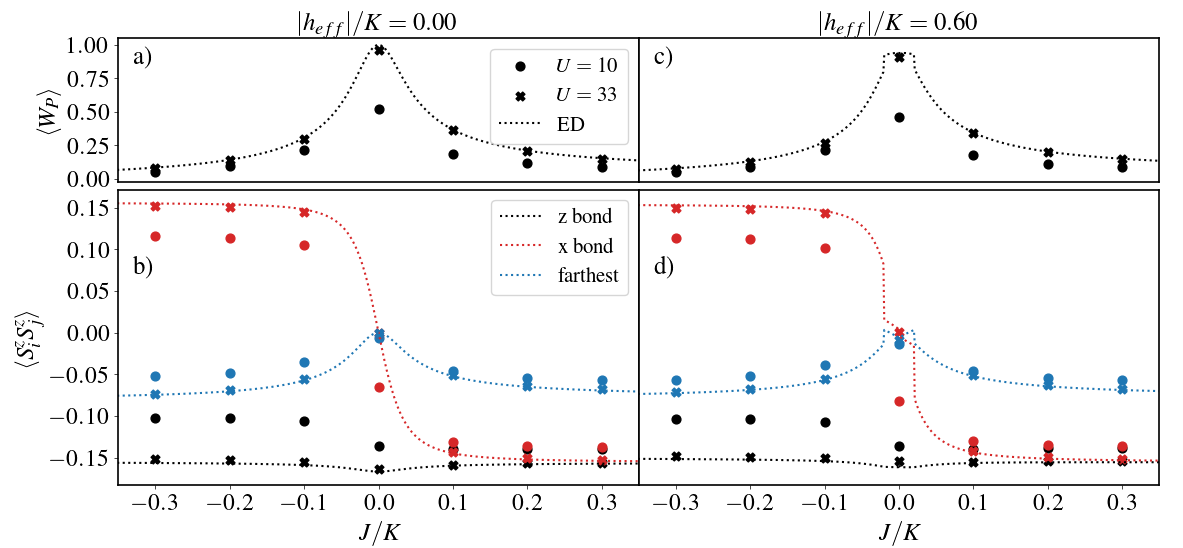}
    \caption{We perform DMRG on twelve Fermi-Hubbard sites at half filling (scatter plot points) as well as exact diagonalization (ED) on six spin-$1/2$ sites using the effective Hamiltonian, Eq.~\eqref{eq:vertex_Ham}.  We set $t_1=h_B=1$, and the values of $t_2$ and $h_V$ are set so as to give a specified value of $J/K$ and $h_\text{eff}$ via Eq.~\eqref{eq:hV_fromeffham}. In (a), (b) $|h_\text{eff}|/K=0$ and in (c),(d) $|h_\text{eff}|/K=0.6$ and the value of $J/K$ is indicated on the horizontal axis. In all plots, the value of $U$ used for DMRG is $U=10$ ($U=33$) for the circle ($\times$) points, respectively. By $U=33$, the DMRG results are almost entirely on top of the ED curves showing that the effective Hamiltonian is a Heisenberg-Kitaev Hamiltonian in a magnetic field. We compare two observables: in (a), (c), we plot the plaquette operator Eq.~\eqref{eq:plaquette_op} and in (c),(d) we plot the correlator $\langle S_i^z S_j^z\rangle$ for a $z$-bond $(i=1,j=3)$, an $x$-bond $(i=1,j=11)$, and the farthest spins $(i=1,j=7)$. In (b) and (d), the color of the points and dashed line indicates which spin correlator is being plotted. A Kitaev plaquette would have $W_P=1$ and the $x$ bond and farthest spin correlators will be zero. Clearly $J/K=0$ and $|h_\text{eff}|/K=0$ satisfy these requirements, but when $|h_\text{eff}|/K\ne 0$, there is a small range of $J/K$ where these are approximately true as well.}
\label{fig:DMRG_ED}
\end{figure*}

In order to verify our theory and clarify possible experimental signatures, we use the Density-Matrix Renormalization Group (DMRG) \cite{White1992} method to directly find the ground state of Eq.~\eqref{eq:12site_Hubb} and compare with exact diagonalization (ED) on six sites given by Eq.~\eqref{eq:vertex_Ham}. For DMRG, we use \texttt{TeNPy} \cite{tenpy} with a bond dimension of $\chi=4096$, large enough to describe the ground state exactly. 

In Fig.~\ref{fig:DMRG_ED}, we plot the plaquette operator, $\langle W_P\rangle $, and  $\langle S_i^z S_j^z\rangle$ for a $z$-bond $(i=1,j=3)$,  an $x$-bond $(i=1,j=11)$, and  the farthest spins $(i=1,j=7)$. We set $t_1=h_B=1$ with $t_2$ and $h_V$ given by
\begin{equation}
\begin{aligned} \label{eq:hV_fromeffham}
h_V &= \frac{-A_1+A_2 \frac{h_\text{eff}}{K}}{A_3 + 4 A_4 \frac{h_\text{eff}}{K}}; \\
t_2 &= \pm \sqrt{U\left(\frac{J}{K}(A_2-4A_4 h_V)-2A_4 h_V-A_5\right)}; \\
A_1 &= 2 \frac{|t_1|^2}{U} \left( 1+ \frac{h_B^2-16|t_1|^2}{4U^2}  + \frac{4|t_1|^2-h_B^2}{2U|h_B|} \right); \\
A_2 &= \frac{2 |t_1|^4}{U^2 |h_B|} + 8 \frac{|t_1|^6}{U^3 h_B^2};\qquad 
A_3 = 1- 2\frac{|t_1|^2}{U^2} \\ A_4 &=\frac{|t_1|^4}{U^2 h_B^2}; \qquad
A_5 = \frac{|t_1|^4}{U^3} - 2\frac{|t_1|^4}{U^2|h_B|} - \frac{4|t_1|^6}{U^3 h_B^2},
\end{aligned}
\end{equation}
so as to reproduce a targeted value of $J/K$ and $|h_\text{eff}|/K$ with errors at higher order than $\mathcal O(U^{-3})$. We are able to verify that DMRG and ED have ground-state energies that agree to $\mathcal O(U^{-4})$ when the parameters are specified in this way (see SM \cite{SM}). We use real values of the hoppings so as to avoid additional parameters needed to compute the flux, $\phi_B$.

We see that when $U/|t_1|=33$, the two methods give excellent agreement with each other further lending credence to our theoretical derivations and plaquette constructions efficacy. Note that, when $|h_\text{eff}|/K=0$ and $J/K\ll 1$, $\langle W_P\rangle$ and $\langle S_i^z S_j^z\rangle$ take on their approximate values as expected for the Kitaev model. Additionally, when $|h_\text{eff}|/K\ne 0$, there are points where the derivatives of these observables are discontinuous, and, in the region $J/K\ll 1$, they take a value close to the Kitaev value. In a large system, these features are consistent with the magnetic field gapping out the itinerant Majoranas and providing a gap that $J$ must overcome; in our system, we have verified that some of the degeneracy seen at the $J/K, |h_\text{eff}|/K=0$ point are lifted in the presence of a magnetic field implying that the same interpretation might hold. 

Taken together, the numerics demonstrate that, even though strictly speaking the Kitaev Hamiltonian only arises at a single point, it is possible to see evidence of the Kitaev state in a range of parameters meaning that the construction is less fine-tuned than anticipated, i.e., there is some robustness.

\section{Conclusion}

In this work, we proposed how to realize the Kitaev honeycomb model, with its non-Abelian anyons, topological order, and its potential as a quantum memory platform with topologically protected quantum information, on connected quantum dots; these small spin-qubit arrays  have already been successfully used as experimental platforms to study many-body collective phenomena such as Mott-Hubbard transitions \cite{hensgens2017quantum,Stafford1994} and Nagaoka ferromagnetism \cite{Nagaoka1966,dehollain2020nagaoka,Buterakos}, making our work both timely and experimentally relevant. 

From the above results (and additional numerical results shown in \cite{SM}), we argue that if $|J|/K \lesssim 0.02$ and $|h_\text{eff}|/K \lesssim 0.75$, our twelve-site system should exhibit Kitaev-spin-liquid-like physics. The experimental setup therefore does not need to be perfectly fine-tuned to observe this physics: for $U/|t_1|=33$ and $|t_1|/|h_B|=1$, these values roughly correspond to a range of $0.0615 \le h_V\le 0.0650$ and $0.2565\le t_2\le 0.2620$ with $K\approx 0.00229$. This range reveals that $h_V$ and $t_2$ need only be accurate at the $5$\% and $2$\% level, respectively, which should be experimentally controllable in semiconductor quantum dot structures. It is also not necessary to prepare the ground-state of the system. As long as the energy is well-below $|h_B|$, the state will always have short-ranged spin-spin correlators as this property is true for every eigenstate in the Kitaev model \cite{Baskaran2007}.

Our proposal, though developed for a single plaquette, works for systems of as many connected plaquettes as is desired. Our expressions can be straightforwardly generalized to include an arbitrary system size or geometry, and we include fully general expressions in the SM \cite{SM} (the only change to Eq.~\eqref{eq:vertex_Ham} is to the coefficient of the $h_V|t_1|^2/U^2$ term). An advantage to having several plaquettes connected to each other is that, for all interior vertex sites (i.e. those sites with a neighbor along an $x$, $y$, and $z$ bond), the applied field on those sites will all be in the $\hat x + \hat y + \hat z$ direction and will therefore be uniform. Alternatively, another advantage of our construction is that the field on the vertex sites does not need to be tuned individually. If the field on the bond sites decays at just the right rate, it can provide the necessary field as $\pmb h_{j\in V}$ points in the same direction as the sum of the neighboring $\pmb h_{j\in B}$. 

The most significant perturbation we have not explicitly included in this letter is a hopping between nearest-neighbor bond sites, $t_3$, that is similar in strength to $t_2$. As is shown in the SM \cite{SM}, the only change to our effective Hamiltonian (besides a constant energy shift) due to this addition is an effective field on the vertex sites at $\mathcal O(U^{-3})$. These terms slightly change the direction of the magnetic field that needs to be applied to a vertex site if the vertex site is not in the interior of the system, but it can be canceled if the field on each vertex site can be tuned individually.

The main experimental difficulty in our proposal is realizing individual magnetic fields on each bond site. In the SM \cite{SM}, we consider what happens if the magnitude and direction of these fields are not perfectly tuned. If the error in the magnitude is $\mathcal O(1/U)$ and if the components of the field of the bond site pointing in the incorrect direction (e.g. in the $\hat x$ and $\hat y$ directions for the $\hat z$ bond) are less than $|t_1|^2/(10U)$, we find some amount of robustness of our chosen signal of spin liquid physics for the ground state, but large enough inaccuracies will wash out the signal. 

Although our proposal should realize a Kitaev spin liquid plaquette in principle, there are many open questions.  For example, what are the necessary conditions and system sizes to observe the non-Abelian anyon braiding signatures?  What are the most suitable experimental signatures of these anyons?  How to realize topological qubits using  anyon braiding in such Kitaev plaquettes?  Our work should motivate both experimental work to realize our proposed Kitaev quantum dot plaquette and theoretical work to answer these questions.

\section{Acknowledgements}
TC is supported by a University of California Presidential Postdoctoral Fellowship and acknowledges support from the Gordon and Betty Moore Foundation through Grant No. GBMF8690 to UCSB. SDS is supported by the Laboratory for Physical Sciences. This research was supported in part by Grant No. NSF PHY-2309135 to the Kavli Institute for Theoretical Physics (KITP). SDS thanks the KITP for its hospitality through the program ``Quantum Materials with and without Quasiparticles.'' Use was made of computational facilities purchased with funds from the National Science Foundation (CNS-1725797) and administered by the Center for Scientific Computing (CSC). The CSC is supported by the California NanoSystems Institute and the Materials Research Science and Engineering Center (MRSEC; NSF DMR 2308708) at UC Santa Barbara.

\clearpage


\section*{Supplementary material}

# Supplemental Material to "Engineering the Kitaev spin-liquid model in a quantum dot system"

Tessa Cookmeyer[1, *] and Sankar Das Sarma[2, 1]

[1]*Kavli Institute for Theoretical Physics, University of California, Santa Barbara, CA, 93106-4030*
[2]*Condensed Matter Theory Center and Joint Quantum Institute, Department
of Physics, University of Maryland, College Park, Maryland 20742-4111*

## I. PERTURBATION THEORY DETAILS

We follow Ref. [S1] but see Ref. [S2] for a different equivalent method. We have a Hamiltonian formally defined by $H = H_0 + \lambda V$, and we have a projection operator $P_0$ that projects onto the $E_0$ energy sector: $H_0 P_0 = P_0 H_0 = E_0 P_0$. We also have a projector $P$ that projects onto the manifold of perturbed states that were originally in the $E = E_0$ manifold. This projector is given by

$$P = P_0 - \sum_{n=1}^{\infty} \lambda^n \sum_{k_1+k_2+\ldots+k_{n+1}=n:k_i \geq 0} S^{k_1} V S^{k_2} V \ldots V S^{k_{n+1}};$$

$$S^k = \left(\frac{1-P_0}{E_0-H_0}\right)^k; \qquad S^0 = -P_0$$

We then define

$$\Gamma = PP_0(P_0 P P_0)^{-1/2}$$

$$(P_0 P P_0)^{-1/2} = P_0 + \sum_{n=1}^{\infty} \frac{(2n-1)!!}{(2n)!!} [P_0(P_0 - P)P_0]^n$$

which leads to an effective Hamiltonian, $H_{\text{eff}} = \Gamma^\dagger H \Gamma$, which we can then determine order-by-order.

Without loss of generality, we pick $E_0 = 0$, and we note that $S^k P_0 = 0$ if $k > 0$, and $SH_0 S = -S$ (since $E_0 = 0$). Using computer algebra software, we arrive at the simplified expression

$$
\begin{aligned}
H_{\text{eff}} = {} & \lambda P_0 V P_0 + \lambda^2 P_0 V S V P_0 + \lambda^3 \left[P_0 V S V S V P_0 - \frac{1}{2}\left(P_0 V S^2 V P_0 V P_0 + P_0 V P_0 V S^2 V P_0\right)\right] \\
& + \lambda^4 \left[P_0 V S V S V S V P_0 + \frac{1}{2}\left(P_0 V S^3 V P_0 V P_0 V P_0 + P_0 V P_0 V P_0 V S^3 V P_0\right)\right. \\
& - \frac{1}{2}(P_0 V P_0 V S V S^2 V P_0 + P_0 V P_0 V S^2 V S V P_0 + P_0 V S V S^2 V P_0 V P_0 + P_0 V S^2 V S V P_0 V P_0 \\
& \left. + P_0 V S^2 V P_0 V S V P_0 + P_0 V S V P_0 V S^2 V P_0)\right] + \ldots,
\end{aligned}
$$

whose first and second terms are clearly just second-order perturbation theory. The fourth term agrees with Ref. [S1] if we specialize, as they did, to the case where $P_0 V P_0 = 0$.

### A. Simplification using the specifics of the Hubbard model

We have

$$H_0 = U \sum_i n_{i\uparrow} n_{i\downarrow}; \qquad \lambda V = \sum_{ij,\sigma} t_{ij} c_{i\sigma}^\dagger c_{j\sigma} + \frac{1}{2} \sum_{i,\sigma,\sigma'} \boldsymbol{h}_i \cdot c_{i\sigma}^\dagger \boldsymbol{\sigma}_{\sigma,\sigma'} c_{i\sigma'} = V_t + V_h$$

where $t_{ij} = t_{ji}^*$ are not necessarily real.

Since $V_h$ does not change the occupancy of sites and $V_t$ does, at half-filling, we have

$$P_0 V_t P_0 = 0; \qquad P_0 V_h S = S V_h P_0 = 0;$$

and for $k > 0$

$$S^k V P_0 = \frac{1}{(-U)^{k-1}} S V P_0; \qquad P_0 V S^k = \frac{1}{(-U)^{k-1}} P_0 V S; \tag{S6}$$

since we can only move to the $E = U$ sector after a single hop. These constraints reduce the number of terms we need to evaluate.

## B. Evaluating terms

As in Ref. [S1], we note the following

$$
\begin{aligned}
P_0 c_{i\uparrow}^\dagger c_{i\uparrow} P_0 &= P_0 c_{i\downarrow} c_{i\downarrow}^\dagger P_0 = \frac{1}{2} + \frac{\sigma^z}{2} \\
P_0 c_{i\downarrow}^\dagger c_{i\downarrow} P_0 &= P_0 c_{i\uparrow} c_{i\uparrow}^\dagger P_0 = \frac{1}{2} - \frac{\sigma^z}{2} \\
P_0 c_{i\downarrow}^\dagger c_{i\uparrow} P_0 &= -P_0 c_{i\uparrow} c_{i\downarrow}^\dagger P_0 = \sigma^- = \frac{1}{2}(\sigma^x - i\sigma^y) \\
P_0 c_{i\uparrow}^\dagger c_{i\downarrow} P_0 &= -P_0 c_{i\downarrow} c_{i\uparrow}^\dagger P_0 = \sigma^+ = \frac{1}{2}(\sigma^x + i\sigma^y).
\end{aligned}
\tag{S7}
$$

Now the terms up to $\lambda^3$ can easily be evaluated

$$
\begin{aligned}
P_0 V_h P_0 &= \frac{1}{2} \sum_{i,\sigma,\sigma'} \boldsymbol{h}_i \cdot \boldsymbol{\sigma}_{\sigma,\sigma'} P_0 c_{i\sigma}^\dagger c_{i\sigma'} P_0 = \frac{1}{2} \sum_i \boldsymbol{h}_i \cdot \boldsymbol{\sigma}_i \\
P_0 V_t S V_t P_0 &= \frac{1}{2} \sum_{ij} \frac{|t_{ij}|^2}{U} (\boldsymbol{\sigma}_i \cdot \boldsymbol{\sigma}_j - 1) \\
P_0 V_t S V_t S V_t P_0 &= \frac{i}{2U^2} \sum_{ijk} t_{ij} t_{jk} t_{ki} \boldsymbol{\sigma}_i \cdot (\boldsymbol{\sigma}_j \times \boldsymbol{\sigma}_k) \\
P_0 V_t S V_h S V_t P_0 &= \frac{1}{4} \sum_{ij} \frac{|t_{ij}|^2}{U^2} (\boldsymbol{h}_j - \boldsymbol{h}_i) \cdot (\boldsymbol{\sigma}_i - \boldsymbol{\sigma}_j) + \frac{1}{2} P_0 V_h P_0 V_t S^2 V_t P_0 + \frac{1}{2} P_0 V_t S^2 V_t P_0 V_h P_0
\end{aligned}
\tag{S8}
$$

For the $\mathcal{O}(\lambda^4)$ terms, we group them by the number of $V_h$ in the expression:

$$
\begin{aligned}
T_0 &= P_0 V_t S V_t S V_t S V_t P_0 + \frac{1}{U} P_0 V_t S V_t P_0 V_t S V_t P_0 \\
T_1 &= P_0 V_t S V_h S V_t S V_t P_0 + P_0 V_t S V_t S V_h S V_t P_0 + \frac{1}{U} (P_0 V_h P_0 V_t S V_t S V_t P_0 + P_0 V_t S V_t S V_t P_0 V_h P_0) \\
T_2 &= P_0 V_t S V_h S V_h S V_t P_0 + \frac{1}{2U^2} (P_0 V_t S V_t P_0 V_h P_0 V_h P_0 + P_0 V_h P_0 V_h P_0 V_t S V_t P_0) \\
&\quad + \frac{1}{U} (P_0 V_h P_0 V_t S V_h S V_t P_0 + P_0 V_t S V_h S V_t P_0 V_h P_0)
\end{aligned}
\tag{S9}
$$

It turns out that $T_1$ is already higher order in a $1/U$ expansion: as noted in the main text, we are considering the ordering $|h_B| \sim t_1 \sim U^{1/4} |h_B|^{1/4} \sqrt{t_2} \sim \sqrt{U |h_V|}$ where $V$ denotes the spins on the vertices and $B$ denotes spins on the bonds. We will also consider a hopping $t_3$ between nearest-neighbor bonds that will be of the same order as $t_2$. The only loops of length three will have a factor $t_1^2 t_2 \sim t_1^3 t_3 \sim t_1^4 / \sqrt{U |h_B|}$, so the term $P_0 V_t S V_t S V_t P_0$, formally at order $\lambda^3$, is already at order $U^{-2.5}$. Therefore, $T_1$ will be $\mathcal{O}(U^{-3.5})$ and not $\mathcal{O}(U^{-3})$ like $T_0$ and $T_2$. These considerations allow us to ignore $T_1$ while carrying out the theory to the desired order consistently.

Since $T_0$ is just the usual term already evaluated in Ref. [S1, S2], we only need to focus on $T_2$. We find

$$P_0 V_t S V_h S V_h S V_t P_0 = -\frac{1}{U}\left\{ P_0 V_t S V_h S V_t P_0 + \frac{1}{4U}(P_0 V_h P_0 V_t S V_t P_0 + P_0 V_t S V_t P_0 V_h P_0), P_0 V_h P_0 \right\} + \frac{1}{4} A_1$$

$$\begin{aligned}
A_1 &= \frac{1}{4}\sum_{ij}\frac{|t_{ij}|^2}{(-U)^3}(\boldsymbol{h}_j - \boldsymbol{h}_i) \cdot \Big( 2\mathrm{Tr}\left[\tilde{M}_j \boldsymbol{\sigma} M_i^T \boldsymbol{\sigma}\right] \cdot (\boldsymbol{h}_j - \boldsymbol{h}_i) \\
&\quad + \left(\mathrm{Tr}\left[\boldsymbol{\sigma} M_i^T \tilde{M}_j \boldsymbol{\sigma}\right] + \mathrm{Tr}\left[\boldsymbol{\sigma}\boldsymbol{\sigma} \tilde{M}_j M_i^T\right]\right) \cdot \boldsymbol{h}_j - \left(\mathrm{Tr}\left[\boldsymbol{\sigma}\boldsymbol{\sigma} M_i^T \tilde{M}_j\right] + \mathrm{Tr}\left[\boldsymbol{\sigma}\tilde{M}_j M_i^T \boldsymbol{\sigma}\right]\right) \cdot \boldsymbol{h}_i \Big) \\
&= \frac{1}{2}\sum_{ij}\frac{|t_{ij}|^2}{(-U)^3}\left[(\boldsymbol{h}_j - \boldsymbol{h}_i)^2 - (\boldsymbol{h}_j - \boldsymbol{h}_i)\cdot\boldsymbol{\sigma}_i\boldsymbol{\sigma}_j\cdot\boldsymbol{h}_j + (\boldsymbol{h}_j - \boldsymbol{h}_i)\cdot\boldsymbol{\sigma}_j\boldsymbol{\sigma}_i\cdot\boldsymbol{h}_i\right]
\end{aligned} \tag{S10}$$

where $\{\cdot,\cdot\}$ is the anticommutator as usual, $\mathrm{Tr}[\cdot]$ denotes the trace, and $M_{i,\alpha\beta} = P_0 c_{i\alpha}^\dagger c_{i\beta} P_0$ and $\tilde{M}_{i,\alpha\beta} = P_0 c_{i\alpha} c_{i\beta}^\dagger P_0$. When substituting into $T_2$, we see that we get

$$T_2 = \frac{1}{4} A_1 + \frac{1}{4U^2}\left[P_0 V_h P_0, [P_0 V_h P_0, P_0 V_t S V_t P_0]\right] \tag{S11}$$

Using the above expressions for $P_0 V_h P_0$ and $P_0 V_t S V_t P_0$ in terms of $\boldsymbol{\sigma}_i$, we can use the commutator relations between the Pauli matrices to evaluate

$$\begin{aligned}
T_2 &= \frac{1}{4} A_1 + \frac{1}{8U^3}\sum_{ij}|t_{ij}|^2\left((\boldsymbol{h}_j - \boldsymbol{h}_i)^2 \boldsymbol{\sigma}_i\cdot\boldsymbol{\sigma}_j - (\boldsymbol{h}_j - \boldsymbol{h}_i)\cdot[\boldsymbol{\sigma}_j\boldsymbol{\sigma}_i\cdot\boldsymbol{h}_j - \boldsymbol{\sigma}_i\boldsymbol{\sigma}_j\cdot\boldsymbol{h}_i]\right) \\
&= \frac{1}{8}\sum_{ij}\frac{|t_{ij}|^2}{U^3}\left((\boldsymbol{h}_j - \boldsymbol{h}_i)^2(\boldsymbol{\sigma}_i\cdot\boldsymbol{\sigma}_j - 1) + (\boldsymbol{h}_j - \boldsymbol{h}_i)\cdot(\boldsymbol{\sigma}_i\boldsymbol{\sigma}_j - \boldsymbol{\sigma}_j\boldsymbol{\sigma}_i)\cdot(\boldsymbol{h}_j + \boldsymbol{h}_i)\right)
\end{aligned} \tag{S12}$$

Therefore, our effective Hamiltonian to $\mathcal{O}(U^{-3})$ is

$$\begin{aligned}
H_{\mathrm{eff}} &= \frac{1}{2}\sum_i \boldsymbol{h}_i\cdot\boldsymbol{\sigma}_i + \frac{1}{2}\sum_{ij}\frac{|t_{ij}|^2}{U}(\boldsymbol{\sigma}_i\cdot\boldsymbol{\sigma}_j - 1) + \frac{i}{2U^2}\sum_{ijk}t_{ij}t_{jk}t_{ki}\boldsymbol{\sigma}_i\cdot(\boldsymbol{\sigma}_j\times\boldsymbol{\sigma}_k) + \frac{1}{4}\sum_{ij}\frac{|t_{ij}|^2}{U^2}(\boldsymbol{h}_j - \boldsymbol{h}_i)\cdot(\boldsymbol{\sigma}_i - \boldsymbol{\sigma}_j) \\
&\quad + T_2 + \sum_{ij}\frac{2|t_{ij}|^4}{U^3}(1 - \boldsymbol{\sigma}_i\cdot\boldsymbol{\sigma}_j) + \frac{1}{2}\sum_{ijk,i\neq k}\frac{|t_{ij}t_{jk}|^2}{U^3}(\boldsymbol{\sigma}_i\cdot\boldsymbol{\sigma}_j - 1) + P_4
\end{aligned} \tag{S13}$$

where $P_4$ is a four-spin term that requires a four-cycle to exist in hopping. The only four-cycle scales as $t_1^2 t_2 t_3$ and is therefore formally at $\mathcal{O}(U^{-4})$ and can be ignored in our case.

## C. Perturbation theory in $|h_B|$

We now want to freeze out the bond spins. The field on the bond spin on an $\alpha$ bond is given by $h_i = -|h_B|\hat{\alpha}$, so that the ground state sector has the spin pointing in the $\hat{\alpha}$ direction. As discussed in the main text, we only have hopping $t_1$ between the vertex and bond spins, $t_2$ between the vertex spins directly, and $t_3$ between the bond spins directly. Note we want $U \gg |h_B|$ for the previous expansion to be valid and $|h_B| \gg |t_1|^2/U$ for perturbation theory in large $h_B$ to be valid.

We still wish to only compute to $\mathcal{O}(1/U^3)$ as before. As discussed in the main text, we ignore the three-spin term because we assume the flux through the triangle formed by the three spins is zero or small. We again split our Hamiltonian

$$\begin{aligned}
H_{\mathrm{eff}} &= H_{0,2} + \lambda V_2; \qquad H_{0,2} = \frac{1}{2}\sum_{i\in B}\left(1 - \sum_{j\in n.n.(i)}\frac{|t_1|^2}{U^2} - \sum_{i'\in n.n.n.(i)}\frac{|t_3|^2}{U^2}\right)\boldsymbol{h}_i\cdot\boldsymbol{\sigma}_i \\
\lambda V_2 &= \frac{1}{2}\sum_{j\in V}\left(1 - \sum_{i\in n.n.(j)}\frac{|t_1|^2}{U^2}\right)\boldsymbol{h}_j\cdot\boldsymbol{\sigma}_j + \sum_{\langle ij\rangle}\left[\frac{|t_1|^2}{U}\left(1 - 4\frac{|t_1|^2}{U^2} + \frac{1}{4}\frac{|h_B|^2}{U^2}\right)(\boldsymbol{\sigma}_i\cdot\boldsymbol{\sigma}_j - 1) + \frac{|t_1|^2}{2U^2}(\boldsymbol{h}_j\cdot\boldsymbol{\sigma}_i + \boldsymbol{h}_i\cdot\boldsymbol{\sigma}_j)\right] \\
&\quad + \sum_{\langle\langle jk\rangle\rangle_V}\left(\frac{|t_2|^2}{U} + \frac{|t_1|^4}{U^3}\right)(\boldsymbol{\sigma}_j\cdot\boldsymbol{\sigma}_k - 1) + \sum_{\langle\langle ii'\rangle\rangle_B}\left[\left(\frac{|t_1|^4}{U^3} + \frac{|t_3|^2}{U}\right)(\boldsymbol{\sigma}_i\cdot\boldsymbol{\sigma}_{i'} - 1) + \frac{|t_3|^2}{2U^2}(\boldsymbol{h}_{i'}\cdot\boldsymbol{\sigma}_i + \boldsymbol{h}_i\cdot\boldsymbol{\sigma}_{i'})\right]
\end{aligned} \tag{S14}$$

where $\langle ij \rangle$ denotes nearest neighbor pairs, $\langle\langle jk \rangle\rangle_{V(B)}$ denotes next-nearest neighbor pairs on vertex (bond) sites, respectively. The notation $n.n(i)$ $[n.n.n.(i)]$ denotes the set of nearest neighbors (next-nearest neighbors) of the site $i$, respectively. As before, we use the scaling $\boldsymbol{h}_{j\in P} \sim t_1^2/U$ and $|t_2|^2 \sim |t_3|^2 \sim |t_1|^4/(U|h_B|)$ to neglect certain higher-order terms.

At leading order in perturbation theory, we have $P_0 V_2 P_0$ (with $\lambda = 1$) where we just replace $\sigma_{i\in B}$ with $\hat\alpha$ if $i$ is on the $\alpha$ bond in the expression for $V_2$ from Eq. (S14). At higher order, we need to identify which terms drive transitions between the low-energy and higher-energy sectors. At lowest order in $U^{-1}$, these terms are

$$V_{t_1} = \frac{|t_1|^2}{U} \sum_{\langle ij\rangle} \boldsymbol{\sigma}_i \cdot \boldsymbol{\sigma}_j; \qquad V_{t_3} = \frac{|t_3|^2}{U} \sum_{\langle\langle ii'\rangle\rangle_B} \boldsymbol{\sigma}_i \cdot \boldsymbol{\sigma}_{i'} \tag{S15}$$

Since there must be at least two of these terms in all higher-order perturbation theory contributions, we can ignore the terms that are already $U^{-3}$. First, we have

$$P_0 V_{t_1} S V_{t_1} P_0 = -2\frac{|t_1|^4}{U^2|h_B|} \left( \sum_{\langle ij\rangle_\alpha} (1 - \sigma_j^\alpha) + \sum_{\langle\langle jk\rangle\rangle_{V,\alpha}} (\boldsymbol{\sigma}_j \cdot \boldsymbol{\sigma}_k - \sigma_j^\alpha \sigma_k^\alpha) \right) \tag{S16}$$

where $\langle ij \rangle_\alpha$ indicates nearest-neighbor pairs with $j \in V$ and $i$ on an $\alpha$ bond, and $\langle\langle jk \rangle\rangle_{V,\alpha}$ indicates a next-nearest-neighbor pair of adjacent vertex sites sharing an $\alpha$ bond.

At order $\mathcal{O}(U^{-3})$, there are two additional terms. Ther first comes from second order in perturbation theory

$$P_0 V_{t_1} S V_{t_3} P_0 + P_0 V_{t_3} S V_{t_1} P_0 = -\frac{|t_1|^2|t_3|^2}{U^2|h_B|} \sum_{\langle ij\rangle_\alpha} \sum_{k_\beta \in n.n.n.(i)} \sigma_i^\beta \tag{S17}$$

where $k_\beta \in n.n.n.(i)$ denotes the set of all next-nearest-neighbors of site $i$ which are on a $\beta$ bond.

The second term arises at third order in perturbation theory from $V_{h_V} = \frac{1}{2} \sum_{i\in V} \boldsymbol{h}_i \cdot \boldsymbol{\sigma}_i + V_{t_1}$ which acts as an effective field. We are most interested in the case when the effective field is at most of the same order as the Kitaev and Heisenberg interactions, which means that $P_0 V_{h_V} P_0 \sim \mathcal{O}(U^{-2})$. Therefore, the only remaining term is

$$P_0 V_{t_1} S V_{h_V} S V_{t_1} P_0 = 2\frac{|t_1|^4}{U^2|h_B|^2} \sum_{\langle\langle jk\rangle\rangle_{V,\alpha}} \left( \frac{1}{2}(h_j^\alpha + h_k^\alpha) - 2\frac{|t_1|^2}{U} \right) \left[ 1 + \boldsymbol{\sigma}_j \cdot \boldsymbol{\sigma}_k - 2\sigma_k^\alpha \sigma_j^\alpha \right] + \mathcal{O}(U^{-4}). \tag{S18}$$

We have not yet specialized to the geometric arrangement of a single plaquette. The above expressions can be used to write the effective Hamiltonian on any number of plaquettes/sites subject to the ordering $|h_B| \sim t_1 \sim U^{1/4}|h_B|^{1/4}\sqrt{t_2} \sim \sqrt{U|h_V|}$ and $t_2 \sim t_3$ and that the effective field on the vertex sites, $P_0 V_{h_V} P_0 \sim U^{-2}$. We finally get the following effective Hamiltonian $H_{V,\text{eff}} = P_0 V_2 P_0 + P_0 V_{t_1} S V_{t_1} P_0 + P_0 V_{t_1} S V_{t_3} P_0 + P_0 V_{t_3} S V_{t_1} P_0 + P_0 V_{t_1} S V_{h_V} S V_{t_1} P_0 + \mathcal{O}(U^{-4})$

$$\begin{aligned}
H_{V,\text{eff}} =& \sum_{j\in V} \left( \frac{1}{2} - \sum_{i_\alpha \in n.n.(j)} \frac{|t_1|^2}{2U^2} \right) \boldsymbol{h}_j \cdot \boldsymbol{\sigma}_j + \sum_{\langle\langle jk\rangle\rangle_{V,\alpha}} \Bigg[ \left( \frac{|t_2|^2}{U} + \frac{|t_1|^4}{U^3} - 2\frac{|t_1|^4}{U^2|h_B|} + \frac{|t_1|^4(h_j^\alpha + h_k^\alpha)}{U^2|h_B|^2} - 4\frac{|t_1|^6}{U^3|h_B|^2} \right) \boldsymbol{\sigma}_j \cdot \boldsymbol{\sigma}_k \\
& + \left( 2\frac{|t_1|^4}{U^2|h_B|} - 2\frac{|t_1|^4(h_j^\alpha + h_k^\alpha)}{U^2|h_B|^2} + 8\frac{|t_1|^6}{U^3|h_B|^2} \right) \sigma_j^\alpha \sigma_k^\alpha \Bigg] \\
& + \sum_{j\in V} \sum_{i_\alpha \in n.n.(j)} \frac{|t_1|^2}{U} \left[ \left( 1 - 4\frac{|t_1|^2}{U^2} + \frac{|h_B|^2}{4U^2} - \frac{|h_B|}{2U} + 2\frac{|t_1|^2}{U|h_B|} \right) \hat\alpha - \frac{|t_3|^2}{U|h_B|} \sum_{k_\beta \in n.n.n.(i)} \hat\beta \right] \cdot \boldsymbol{\sigma}_j + C \\
C =& -3|h_B| \left( 1 - \frac{2|t_1|^2}{U^2} - \frac{2|t_3|^2}{U^2} \right) - 12\frac{|t_1|^2}{U} \left( 1 - 4\frac{|t_1|^2}{U^2} + \frac{1}{4}\frac{|h_B|^2}{U^2} \right) - 6\left( \frac{|t_2|^2 + |t_3|^2}{U} + 2\frac{|t_1|^4}{U^3} \right) \\
& - 24\frac{|t_1|^4}{U^2|h_B|} - 24\frac{|t_1|^6}{U^3|h_B|^2} + \sum_j \sum_{i_\alpha \in n.n.(j)} \left( \frac{|t_1|^2}{2U^2} + \frac{|t_1|^4}{U^2|h_B|^2} \right) h_j^\alpha
\end{aligned} \tag{S19}$$

where the constant, $C$, is specialized to the single plaquette we are considering but the other terms in this expression are fully general. Similar to above, the notation $i_\alpha \in n.n.(j)$ indicates nearest-neighboring sites to site $j$ with bond label $\alpha$.

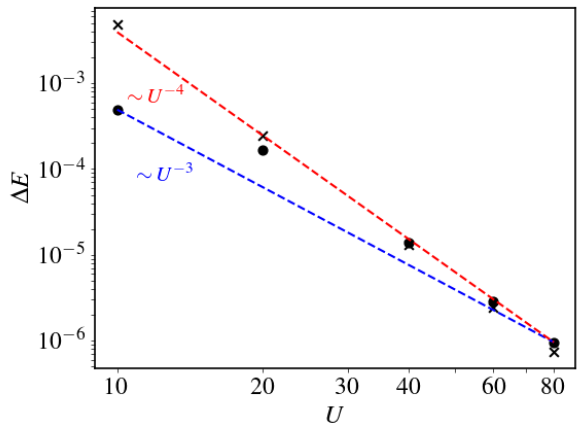

Figure S1. We perform DMRG on twelve Fermi-Hubbard sites, DMRG on the effective spin Hamiltonian on twelve sites using Eq. (S14), and exact diagonalization (ED) on six sites using the effective Hamiltonian, Eq. (S26). We set $3t_1 = 2h_B = 3$, and the values of $|t_2|$ and $h_V$ are set so as to give $J/K = 0.3$ and $h_{\text{eff}} = 0.05$ using Eq. (S21) (with $t_3$ set to zero) and then we set $t_3 = 0.9t_2$. We plot the difference in energies between the Hubbard model and the twelve-site spin model (circles) and the different between the twelve-site spin model and the ED (x's) as a function of $U$ and include dashed lines to indicate the scaling $U^{-4}$ in red and $U^{-3}$ in blue. We see that difference between all three methods are converging as $U^{-4}$, and we are neglecting terms at that order so it should not be faster.

Note that if we have several plaquettes, the effective magnetic field for an interior site $j$ (i.e. a site having nearest neighbors along $x$, $y$, and $z$ bonds) is given by

$$h_{V,\text{eff},\text{int},j} = \left(\frac{1}{2} - \frac{3|t_1|^2}{2U^2}\right)\boldsymbol{h}_j + \frac{|t_1|^2}{U}\left(1 - 4\frac{|t_1|^2}{U^2} + \frac{|h_B|^2}{4U^2} - \frac{|h_B|}{2U} + 2\frac{|t_1|^2}{U|h_B|} - \frac{2|t_3|^2}{U|h_B|}\right)(\hat{x} + \hat{y} + \hat{z}), \quad \text{(S20)}$$

which means that a uniform magnetic field applied to the system could cancel the effective magnetic fields on all interior vertex sites (but stronger non-uniform fields still need to be applied on the interior bond sites).

For our single plaquette, notice that the only effect of $t_3 \sim t_2$ besides a constant shift in the energy is to provide an $\mathcal{O}(U^{-3})$ correction to the effective field. In addition to being able to cancel this term with a correctly applied magnetic field on the vertex sites, the term becomes negligible as $U$ becomes large since the Kitaev term $K \sim U^{-2}$. For simplicity of the analysis, we will therefore set $t_3 = 0$. Additionally, although the effect of $t_3$ on the numerics we presented in the main text may produce a noticeable quantitative effect, we find that a small magnetic field actually helps stabilize the Kitaev spin liquid physics in our small system and thus our qualitative results, namely that our setup allows for the creation of the Kitaev spin-liquid on a single plaquette, is not affected by nonzero $t_3$.

We can now tune $\boldsymbol{h}_{j\in V}$ and $|t_2|$ to get an effective Heisenberg-Kitaev Hamiltonian in a magnetic field. It is easy to see that in order for the Hamiltonian to be close the the Kitaev Hamiltonian, we need $|\boldsymbol{h}_j| \sim |t_1|^2/U$ and $|t_2|^2/U \sim |t_1|^4/(U^2|h_B|)$ justifying our use of these relations earlier. If we want to instead target a particular $J/K$ and $\boldsymbol{h}_{\text{eff}}/K$, we can easily work out the corresponding values of $|t_2|$ and $h_V$. Namely, if $\boldsymbol{h}_{\text{eff}}$ is parallel to $\boldsymbol{h}_j = |h_V|(\hat{\alpha} + \hat{\beta})$ where $\alpha$ and $\beta$ are the two bonds connected to the vertex $j$, as quoted in the main text, we get the general expressions

$$
\begin{aligned}
&h_V = \frac{-A_1 + A_2\frac{h_{\text{eff}}}{K}}{A_3 + 4A_4\frac{h_{\text{eff}}}{K}}; \qquad t_2 = \sqrt{U\left(\frac{J}{K}(A_2 - 4A_4h_V) - 2A_4h_V - A_5\right)}; \\
&A_1 = 2\frac{|t_1|^2}{U}\left(1 - 4\frac{|t_1|^2}{U^2} + \frac{h_B^2}{4U^2} - \frac{|h_B|}{2U} + 2\frac{|t_1|^2}{U|h_B|}\right); \qquad A_2 = \frac{2|t_1|^4}{U^2|h_B|} + 8\frac{|t_1|^6}{U^3h_B^2} \\
&A_3 = 1 - 2\frac{|t_1|^2}{U^2}; \qquad A_4 = \frac{|t_1|^4}{U^2h_B^2}; \qquad A_5 = \frac{|t_1|^4}{U^3} - 2\frac{|t_1|^4}{U^2|h_B|} - \frac{4|t_1|^6}{U^3h_B^2}.
\end{aligned}
\quad \text{(S21)}
$$

## II. NUMERICAL RESULTS

In this section, we will evaluate Eq. (S4) for the twelve sites on our single hexagon plaquette with the Density-Matrix Renormalization group (DMRG) method, Eq. (S14) for the twelve site spin-Hamiltonian using DMRG, and

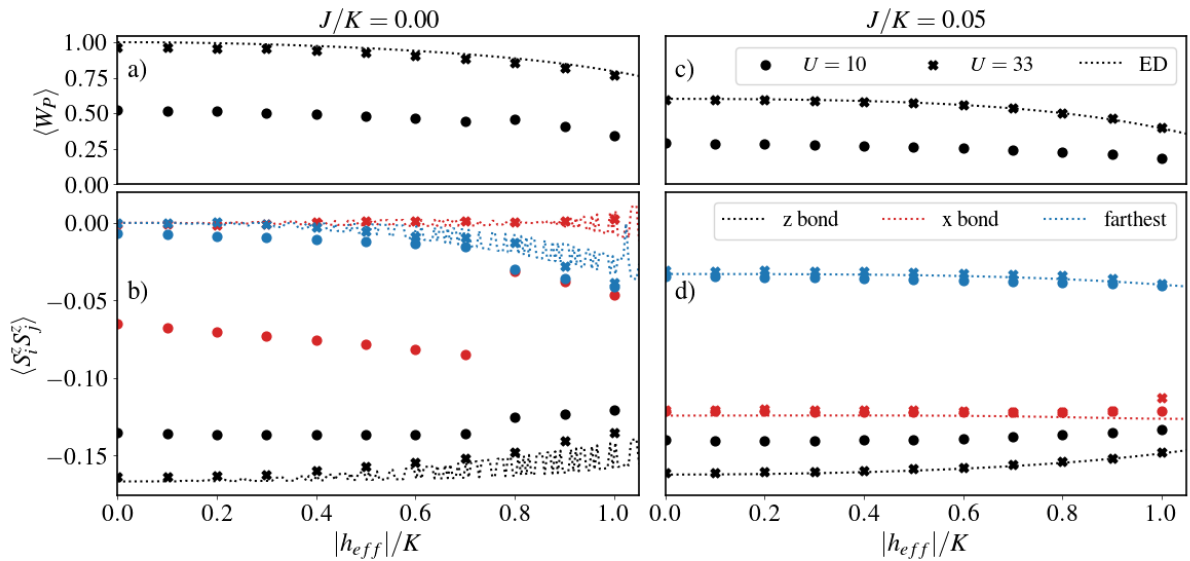

Figure S2. We perform DMRG on twelve Fermi-Hubbard sites (scatter plot points) as well as exact diagonalization (ED) on six sites using the effective Hamiltonian, Eq. (S26). We set $t_1 = h_B = 1$, and the values of $|t_2|$ and $h_V$ are set so as to give a specified value of $J/K$ and $h_{eff}$ via Eq. (S21) (and $t_3 = 0$). In (a), (b) $J/K = 0$ and in (c),(d) $J/K = 0.05$ and the value of $|h_{eff}|/K$ is indicated on the horizontal axis. In all plots, the value of $U$ needed for DMRG is $U = 10$ for the circle points and $U = 33$ for the × points. By $U = 33$, the DMRG results are almost entirely on top of the ED curves showing that the effective Hamiltonian is a Heisenberg-Kitaev Hamiltonian in a magnetic field. We compare two observables: in (a), (c), we plot the plaquette operator $\langle W_P \rangle$ [see Eq. (4) from the main text] and in (c),(d) we plot the correlator $\langle S_i^z S_j^z \rangle$ for a $z$-bond ($i = 1, j = 3$), an $x$-bond ($i = 1, j = 11$), and the farthest spins ($i = 1, j = 7$). In (b) and (d), the color of the points and dashed line indicates which spin correlator is being plotted. A Kiteav plaquette would have $W_P = 1$ and the $x$ bond and farthest spin correlators will be zero. Clearly $J/K = 0$ and $|h_{eff}|/K = 0$ satisfy these requirements, but when $|h_{eff}|/K \neq 0$, there is a range of values where these restrictions are approximately satisfied. Notably, the magnetic field is a much weaker perturbation that $J/K$ (see Fig. 2 in the main text).

Eq. (S26) with exact diagonalization on the six vertex spins. We carry out the DMRG calculations using TeNPy [S3]. To specify parameters, we pick a value of $|t_1|$, $|h_B|$. We then fix a value of $U$, and then set $|t_2|$ and $h_V$ to reproduce a particular value of $J/K$ and $|h_{eff}|/K$. Because we use the maximum possible bond dimension of 4096 for our twelve Fermi-Hubbard sites, the remaining parameters are likely not important, but for completeness we list them here: we check for convergence every two sweeps and use a mixer for the first two sweeps. Our energy and entropy cutoffs to ensure convergence are $3 \times 10^{-8}$ and $3 \times 10^{-5}$ respectively.

First, as shown in Fig. S1, we verify that the difference in energy between the three methods decreases as $U^{-4}$ indicating that our perturbation theory results are accurate to the order we expect. We set the value of $t_2$ and $h_V$ using Eq. (S21) to reproduce a particular value of $h_{eff}/K$ and $J/K$, and we set $t_3 = 0.9t_2$. Although having non-zero $t_3$ would change the equation Eq. (S21), we ignore its effect for the purpose of Fig. S1 (and we set $t_3 = 0$ for all other calculations). Note that if there is significant non-zero flux through the three cycles composed of two vertices and a bond, we will be making errors at order $U^{-3.5}$ since we neglected the term $T_1$ above.

From now on, we will only compare DMRG on the twelvec Fermi-Hubbard model and ED on the six vertex spins and $|t_3| = 0$. We set $|t_1| = |h_B| = 1.0$ and $U = 10$ or $U = 33$. In Fig. 2 in the main text, we use fix values of $|h_{eff}|/K$ and change $J/K$. In Fig. S2, we fix a value of $J/K$ and change $|h_{eff}|/K$. As we can see, we obtain excellent agreement when $U = 33$. It is worth noting that the ground state is degenerate when $J/K = 0$, leading to a range of possible values of the spin-spin correlators (all degenerate states have the same value of $\langle W_P \rangle$), which is why the ED curves jump around very quickly as a function of parameters. A small $J/K$ lifts this degeneracy.

From both of these figures, we see that there is a large range in $|h_{eff}|/K$ and a small range in $J/K$ where we have $\langle W_P \rangle \approx 1$ and $\langle S_i^z S_j^z \rangle$ short-ranged, which we will refer to as being in a spin-liquid plaquette regime. Because we have good agreement between ED and DMRG when $U = 33$, we can explore a wider set of phase space just using the much easier (and faster) to use ED technique. In Fig. S3, we plot $\langle W_P \rangle$ and $\langle S_i^z S_{11}^z \rangle$, the $S^z$-$S^z$ spin-correlator on an $x$ bond, for a wide range of $J/K$ and $|h_{eff}|/K$. We can see that the spin-liquid plaquette regime exists in approximately the range $|J/K| \leq 0.01 - 0.02$ and $|h_{eff}|/K \leq 0.75$. It is also clear from our plots that there appear to be first-order transitions out of this regime when $|h_{eff}|/K \neq 0$. For a large system, the isotropic Kitaev model is known to be gapless, but the gap can be lifted by a magnetic field [S4]. Therefore, we would expect the spin-liquid

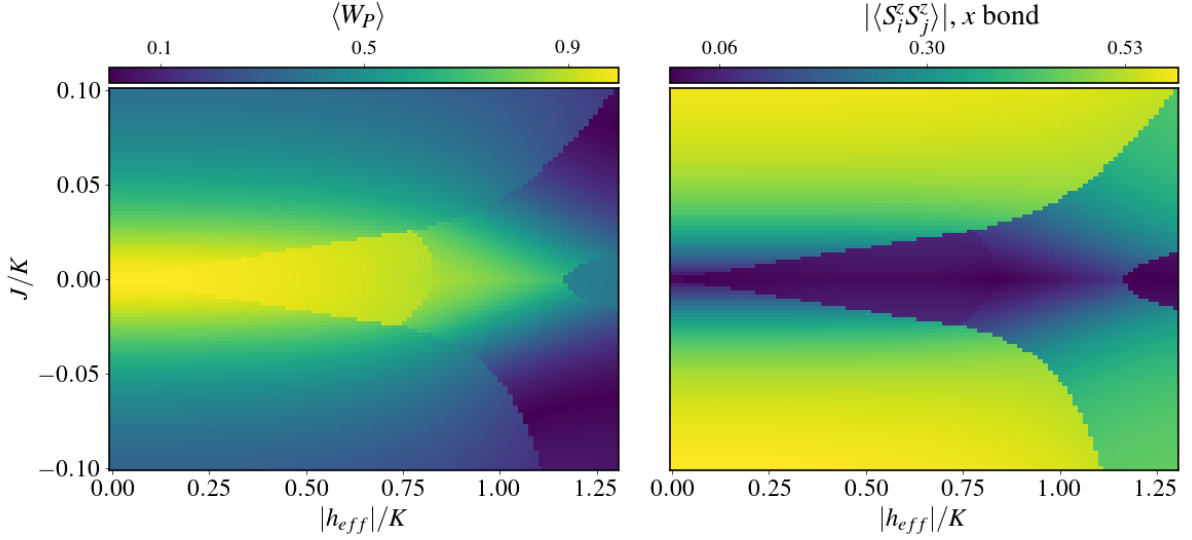

Figure S3. Using ED, we plot $\langle W_P \rangle$ and $\langle S_1^z S_{11}^z \rangle$, the correlator on an $x$-bond for a range of values $J/K$ and $|h_{\mathrm{eff}}|/K$. We can see that $\langle W_P \rangle \approx 1$ and $\langle S_1^z S_{11}^z \rangle \approx 0$ (their values at the Kitaev point) in a range of parameters. We also see that there are sharp changes in the plotted values once $|h_{\mathrm{eff}}|/K \neq 0$. We interpret these features as due to the small-system analog of the magnetic field opening a gap: when $J/K = 0$, the field lifts some of the degeneracy seen at $|h_{\mathrm{eff}}|/K = 0$ but not all. Nevertheless, it seems to open up a range of values where the Kitaev spin-liquid ground state on a single plaquette can be observed.

state to be more robust to perturbations in a small magnetic field. Although $|h_{\mathrm{eff}}|/K$ does not lift all the degeneracy seen at $J/K = |h_{\mathrm{eff}}|/K = 0$, it does lift some of the degeneracy, and this lifting may have to be overcome by $J/K$ to exit the spin-liquid plaquette regime. We therefore interpret the sharp features in Fig. 2 as the analog to gapping out the Kitaev spin-liquid in a magnetic field.

## III. IMPERFECTIONS IN THE APPLIED MAGNETIC FIELDS

In deriving Eq. (S26), we assume that the field strength on each of the bonds was the same. Since the control of site specific magnetic field is likely to be the most difficult aspect of our construction, it is worth analyzing what occurs when the field strength is not constant and the field direction deviates from the precise direction we have assumed. We will assume that the field strength on each of the bond sites is within some range $|h_B| - \delta h_B < h_{i_\alpha}^\alpha < |h_B| + \delta h_B$ and there is some deviation from the correct field direction, $\boldsymbol{h}_{i_\alpha} \cdot \hat{\beta} = \delta v_i^\beta \ll |h_B|$. We will assume that $\delta h_B \lesssim 1/U$ and $\delta v_i^\beta \lesssim 1/U$ and only perform this calculation to $\mathcal{O}(U^{-2})$. The imperfection in the applied field on the vertex sites will be included below.

Our calculation remains the same up to Eq. (S14) except that we now change

$$H_{0,2} = \frac{1}{2} \sum_{i_\alpha \in B} \sigma_i^\alpha h_i^\alpha \left( 1 - \frac{2|t_1|^2}{U^2} - \frac{2|t_3|^2}{U^2} \right) \tag{S22}$$

and add a term to $\lambda V_2$ that is

$$\delta V_2 = \frac{1}{2} \sum_{i_\alpha \in B} \sum_{\beta \neq \alpha} \delta v_i^\beta \sigma_i^\beta. \tag{S23}$$

In perturbation theory, we will have two additional terms arising from

$$P_0 \delta V_2 S \delta V_2 P_0 = -\frac{1}{4} \sum_{i_\alpha \in B} \sum_{\beta \neq \alpha} \frac{(\delta v_i^\beta)^2}{|h_i^\alpha|} \tag{S24}$$

and

$$P_0 \delta V_2 S V_{t_1} P_0 + P_0 V_{t_1} S \delta V_2 P_0 = -\frac{|t_1|^2}{U} \sum_{\langle ij \rangle_\alpha} \frac{\boldsymbol{h}_i - h_i^\alpha \hat{\alpha}}{|h_i^\alpha|} \cdot \boldsymbol{\sigma}_j \tag{S25}$$

The original perturbation theory contribution, Eq. (S16), only needs to be modified to have $|h_{i_\alpha}^\alpha|$ in the denominator instead of $|h_B|$ since the denominator arises from which bond site's direction is flipped by each term in $V_{t_1}$. Our new effective Hamiltonian $H'_{V,\text{eff}}$ (up to a constant) is then

$$H'_{V,\text{eff}} = \sum_{j \in V} \left[ \frac{\boldsymbol{h}_j}{2} + \sum_{i_\alpha \in n.n.(j)} \frac{|t_1|^2}{U} \left[ 2 \frac{|t_1|^2}{U|h_{i_\alpha}^\alpha|} \hat{\alpha} + \left( \frac{1}{2U} - \frac{1}{|h_i^\alpha|} \right) \boldsymbol{h}_i \right] \right] \cdot \boldsymbol{\sigma}_j$$
$$+ \sum_{\langle\langle jk \rangle\rangle_{V,\alpha}} \left[ \left( \frac{|t_2|^2}{U} - 2 \frac{|t_1|^4}{U^2|h_{i_\alpha}^\alpha|} \right) \boldsymbol{\sigma}_j \cdot \boldsymbol{\sigma}_k + \left( 2 \frac{|t_1|^4}{U^2|h_{i_\alpha}^\alpha|} \right) \sigma_j^\alpha \sigma_k^\alpha \right]$$

(S26)

where we used $h_i^\alpha = -|h_i^\alpha|$ and, in the sum over $\langle\langle jk \rangle\rangle_{V,\alpha}$, the site $i_\alpha$ is the intermediate bond site between the two vertex sites. We can now see that variations in the strength and direction of the magnetic field on the bond sites will produce bond-dependent Heisenberg interactions $J_{\langle ij \rangle}$, Kitaev interactions $K_{\langle ij \rangle}$, and magnetic fields $\boldsymbol{h}_j$. As before, $K = 2|t_1|^4/(U^2|h_B|)$, and $\delta K/K, \delta J/K \lesssim \delta h/|h_B| \sim \mathcal{O}(U^{-1})$, where $|K_{\langle ij \rangle} - K| \lesssim \delta K$ and similarly for $J$. For the magnetic field, there are two contributions—one due to $\delta h$ and the other due $\delta v_i^\beta$. The former has $\delta \boldsymbol{h}_{\text{eff},1}/K \lesssim \delta h/|h_B| \sim \mathcal{O}(U^{-1})$ and the latter has $\delta \boldsymbol{h}_{\text{eff},2}/K \lesssim \delta v_i^\beta/(|t_1|^2/U) \sim \mathcal{O}(U^0)$. Additionally, there is the error we expect in the applied field on the vertex sites; if the error is $\mathcal{O}(U^{-2})$ (since $\boldsymbol{h}_{j\in V} \sim 1/U$), it will also contribute an error $\delta \boldsymbol{h}_{\text{eff},3}/K \lesssim \mathcal{O}(U^0)$, but we will absorb this contribution into the definition of $\delta \boldsymbol{h}_{\text{eff},2}$.

The most severe effect, then, is that the effective field direction can be shifted away from its ideal direction. In any given experimental setup, there would be a single realization of these imperfections. We can simulate these conditions by picking a random distribution of $J_{\langle jk \rangle}$, $K_{\langle jk \rangle}$ and adding a random component to the effective field on the vertex sites $\boldsymbol{h}_i = |h_{\text{eff}}|(\hat{\alpha} + \hat{\beta}) + \delta v \boldsymbol{v}$ (where site $j$ borders and $\alpha$ and $\beta$ bond). Note that the effective Kitaev coupling on a given bond is $K_{\langle jk \rangle} = K + \delta K_{\langle jk \rangle}$ and that the effective Heisenberg coupling $J_{\langle jk \rangle} = J + \delta J_{\langle jk \rangle}$ share the same anisotropy: $\delta K_{\langle jk \rangle} = -\delta J_{\langle jk \rangle}$. The values $\delta K_{\langle jk \rangle} \in [-0.03, 0.03]$ are uniformly distributed to imitate the range expected for $U = 33$. To the effective field found in the ideal case, we add a field $\delta v \boldsymbol{v}$ where $\boldsymbol{v}$ is uniformly distributed across the unit ball to capture the independent source of randomness of stray fields and the imprecision in the fields applied to the vertex sites. We tune $\delta v$ independently to get a sense of the effect of the size of these field imperfections. These approximations likely do not perfectly replicate the variance expected but are meant to give an estimate of their effect.

In Fig. S4 we remake the plot from Fig. S3 for three different realizations of the distributions described in the previous paragraph. When $\delta v = 0$ (the first row), we see that the presence of proximity to the Kitaev point is still detectable. The strength of the signal decreases quite significantly as $\delta v$ is increased, but $\delta v \lesssim 0.1$ still appears to have some signal. In total, if the magnitude of the field $|h_B|$ is accurate at the percent level and if the strength of the other components of the magnetic field, $\delta v_i^\beta \lesssim |t_1|^2/(10U)$, then there will be detectable remnants of the Kitaev spin liquid.

---

* tcookmeyer@kitp.ucsb.edu

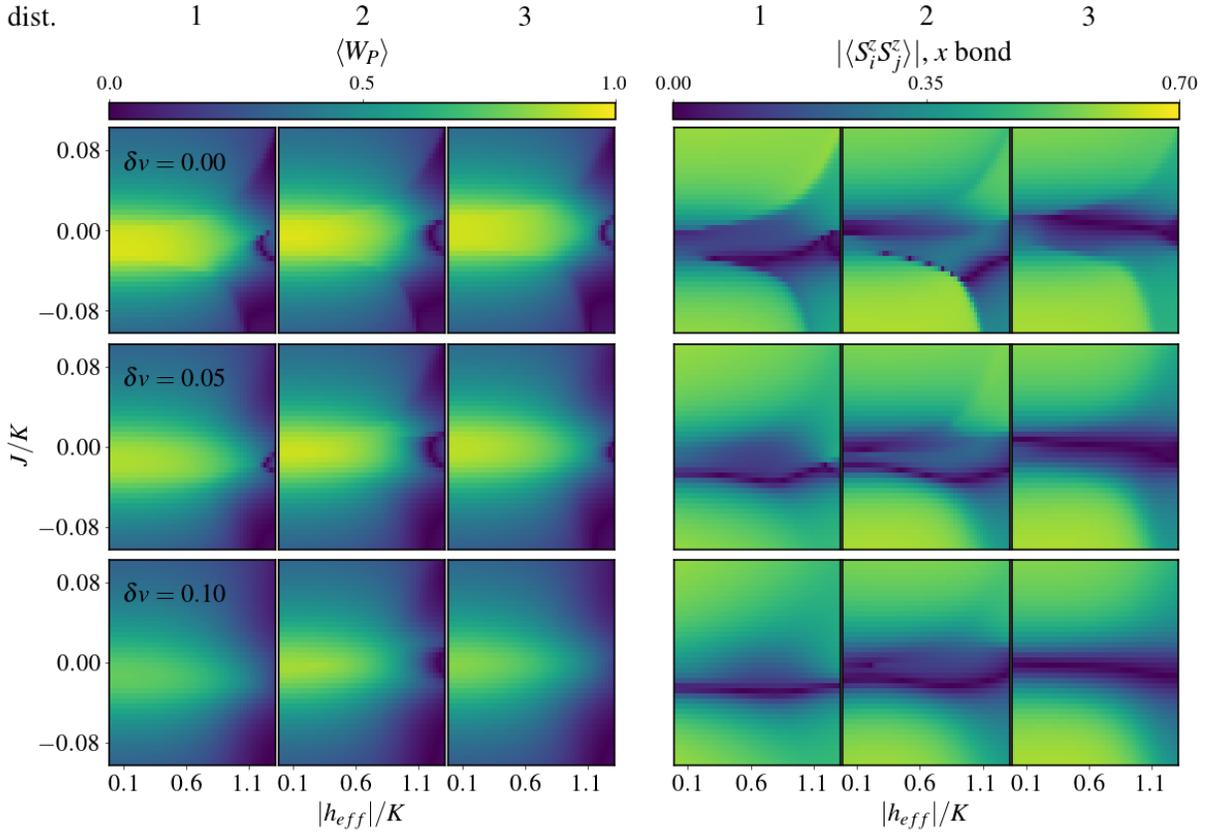

Figure S4. Using ED, we plot $\langle W_P \rangle$ and $\langle S_1^z S_{11}^z \rangle$, the correlator on an $x$-bond for a range of values $J/K$ and $|h_{\text{eff}}|/K$ where we have introduced randomness in the strength of the magnetic field $\boldsymbol{h}_{i_\alpha}^\alpha = |h_B| + \delta h_B$ on each bond site and included a random effective field on each vertex $\boldsymbol{h}_j = |h_{\text{eff}}|(\hat\alpha + \hat\beta) + \delta v \boldsymbol{v}$ (where site $j$ borders an $\alpha$ and $\beta$ bond). The random field strength components $\delta h_B \in [-0.03, 0.03]$ is uniformly distributed and the random field direction $\boldsymbol{v}$ is uniformly distributed over the unit ball. As argued in the text, this choice of randomness should roughly capture the effect of applying imperfect magnetic fields on the twelve Hubbard sites. We makes these plots for three distributions in total where the $n$th column on each side of the figure corresponds to the $n$th distribution, as indicated at the top of the figure. These plots should be compared with Fig. S3. When $\delta v = 0$, we see that the randomness in the field strength does not change the signal too significantly and the broad features (i.e. $\langle W_P \rangle \approx 1$ and $\langle S_i^z S_j^z \rangle \approx 0$) can still be observed. These features are much more sensitive to $\delta v$, but there still remain remnants of the broad features.



\end{document}